\documentclass[preprint2]{aastex}

\newcommand{\CB} {\textnormal{COMBO-17}}
\newcommand{\C} {\textnormal{COMBO-17+4}}
\newcommand{\OM} {\emph{\textrm{OMEGA2k}}}
\newcommand{\MPIAPHOT} {\textit{MPIAPHOT}}

\slugcomment{Not to appear in Nonlearned J., 45.}

\shorttitle{Red-Sequence Galaxies at High Redshift}
\shortauthors{Nicol et al.}

\begin{document}


\title{Red-Sequence Galaxies at High Redshift by the COMBO-17+4 Survey}


\author{Marie-H{\'e}l{\`e}ne Nicol\altaffilmark{1},  Klaus Meisenheimer\altaffilmark{1}, Christian Wolf\altaffilmark{1,2} and Christian Tapken\altaffilmark{3}}
\altaffiltext{1}{Max-Planck Institut f{\"u}r Astronomie, K{\"o}nigstuhl 17,  D-69117 Heidelberg, Germany}
\email{nicol@mpia.de, meise@mpia.de}
\altaffiltext{2}{Department of Physics, Denys Wilkinson Building, University of Oxford, Keble Road, Oxford OX1 3RH, UK}
\email{cwolf@astro.ox.ac.uk}
\altaffiltext{3}{Astrophysikalisches Institut Potsdam, An der Sternwarte 16, D-14482 Potsdam, Germany}
\email{ctapken@aip.de}


\begin{abstract}
We investigate the evolution of the galaxy population since redshift 2 with a focus on the colour bimodality and mass density of the red sequence.
We obtain precise and reliable photometric redshifts up to $z=2$ by supplementing the optical survey COMBO-17 with observations in four near-infrared bands on 
0.2 square degrees of the COMBO-17 A901-field. Our results are based on an $H$-band-selected catalogue of 10692 galaxies complete to $H=21\fm 7$.
We measure the rest-frame colour $(U_{280}-V)$ of each galaxy, which across the redshift range of our interest requires no extrapolation and is robust
against moderate redshift errors by staying clear of the 4000\AA -break.
We measure the colour-magnitude relation of the red sequence as a function of lookback time from the peak in a colour error-weighted histogram, and thus 
trace the galaxy bimodality out to $z\simeq 1.65$. The $(U_{280}-V)$ of the red sequence is found to evolve almost linearly with lookback time. 
At high redshift, we find massive galaxies in both the red and the blue population. Red-sequence galaxies with $\log M_*/M_\sun>11$ increase in mass density by 
a factor of $\sim 4$ from $z\sim 2$ to 1 and remain nearly constant at $z<1$. However, some galaxies as massive as $\log M_*/M_\sun=11.5$ are already in place at $z\sim2$. 
\end{abstract}





\section{INTRODUCTION}

The Lambda Cold Dark Matter ($\Lambda$CDM) cosmological model predicts a bottom-up hierarchical scenario of structure formation in which large
structures have been formed recently ($z\lesssim1$) through mergers of existing smaller structures. It is still unclear when and how in this 
context galaxies have formed and how they have evolved into the baryonic structures made of stars we see today. The most evolved and massive 
systems known in the Universe are found especially among red-sequence galaxies. They are straightforward to identify and seem to mark an end 
point in galaxy formation. Model predictions can thus be tested against measurements of the build-up of this population through cosmic time.
To this end, the population needs to be tracked towards higher redshift. Their colour, mass and number density hold clues about 
their mass assembly over time. 

It is still unknown when in the history of the universe the red-sequence population first emerged.
Using the optical \CB~ data, \cite{bell04} have shown that the galaxy bimodality is present at all redshifts out to $z=1$ and that the
red sequence was already in place at $z=1$, see also \cite{weiner05}. Several spectroscopic studies have observed massive red galaxies at 
$z\gtrsim1.5$ \citep{cimatti04,glazebrook04,daddi05,mcgrath07,conselice07,kriek08,cassata08}, but did not allow the study of large samples. 
On the other hand, photometric redshift surveys combining optical and NIR data can provide photometric redshifts for larger samples of high redshift galaxies. 
Recently, \cite{taylor09} demonstrated the persistence of the red sequence up to $z\sim1.3$ and \cite{cirasuolo07} as well as \cite{franzetti07} up to $z\sim1.5$. 
Using colour-colour diagrams \citep{williams09} claimed the presence of quiescent galaxies analogous to a red sequence up to $z\sim2$.
A persistent issue for photometric redshift surveys is an increasing uncertainty in the measurement of rest-frame colours at high redshift ($z\gtrsim1.5$). 
It increases the scatter in colour-magnitude and colour-stellar mass diagrams,
thus blurring the signature of a possible red sequence at high
redshift. Especially, the expected small red-sequence population in field
samples may be undetectable.

In this paper, we maximise the separation of red quiescent and blue star-forming galaxies by using the rest-frame colour $(U_{280}-V)$ as a diagnostic. 
\citet{wolf09} have shown how in galaxy clusters this colour also tends to render star-forming red galaxies bluer to leave a more purified red sample.
We use colour-error-weighted histograms to improve the recognition of a red sequence in the face of increasing colour uncertainties at high redshift
(see Sect.~3 for all methods). 
In Sect.~4 we investigate the colour distribution up to $z=2$ and derive a colour-magnitude relation and a bimodality separation evolving with redshift.
Thus, we attempt to estimate when the red-sequence population emerges. Ultimately, we quantify the evolution of the following galaxy properties: 
colour, luminosity, mass and number density for the red-sequence galaxy population since $z=2$ (Sect.~5).
We summarize and conclude this study in Sect.~6

Throughout this paper, we use the cosmological parameters $\Omega_m=0.3$, $\Omega_{\Lambda}=0.7$ and $H_0=70.7$ km s$^{-1}$ Mpc$^{-1}$. 
All magnitudes are in the Vega system.

\section{OBSERVATIONS}

\subsection{The COMBO-17+4 Survey}

The \C~ survey is the NIR extension of the optical \CB~ survey and is designed to probe galaxy evolution since $z=2$. Near infrared data were necessary
to obtain optical rest-frame properties for galaxies in the redshift range $1<z<2$, where the optical rest-frame of galaxies is shifted into the NIR. 
The survey consists of observations in four NIR bands $(\lambda/\Delta\lambda)$: the three medium-bands $Y(1034/80)$, $J_1(1190/130)$, $J_2(1320/130)$,
and the broad-band $H(1650/300)$. In the long run, the \C~ survey targets three independent fields (A901, A226, and S11) for a total coverage of 0.7$\sq\degr$
(see Table~\ref{tbl-1} for coordinates and integration times). The results derived in this paper are based on the observations of the A901-field only. 
This field contains the super-cluster of galaxies A901/2 located at $z=0.165$ where further multi-wavelength coverage from X-ray to radio was obtained 
by the STAGES survey \citep{gray09}.

\subsection{Data}

The NIR data were obtained in several observing runs from December 2005 to April 2009 with the NIR wide field camera Omega2000 at the prime focus of the 
3.5-m telescope at Calar Alto Observatory in Spain. The camera has a pixel size of $0.45\arcsec$ and a wide field of view of $15.4\arcmin\times 15.4\arcmin$,
so that a half-degree COMBO-17 field can be covered with a $2\times 2$-mosaic. On the A901-field, only three of four pointings could be finished in the time 
awarded to the project, and hence the NIR data are missing in the south-west quadrant. 

Also, an area of $68\arcsec\times 76\arcsec$ centred at $(\alpha,\delta)_{J2000}=09^h 56^m 32^s.4,-10\degr 01\arcmin 15\arcsec$ was cut out 
to avoid spurious objects created by the halo of a very bright ($K=5\fm75$) Mira variable star, also known as the IRAS point source 09540-0946. The total NIR 
coverage is thus 690$\sq\arcmin \simeq 0.19\sq\degr$. The optical data include a combination of 17 broad and medium bands centred between 365\,nm and 915\,nm and 
were obtained with the Wide Field Imager (WFI) at La Silla Observatory between February 1999 and January 2001 by the \CB~ survey \citep{wolf03, gray09}.

\subsection{Data Analysis}

The NIR data reduction and photometry were performed using the software ESO-MIDAS in combination with the \MPIAPHOT~ package developed by \cite{RM91}
and the \OM~ data reduction pipeline developed by \cite{fassbenderTH}. 
The image data reduction consisted of flatfielding, dark current and sky background subtraction as well as correction for bad pixels and
pixel hits by cosmic ray events. Additive stray light appearing in a ring shape in all scientific and calibration images taken with the
Omega2000 camera have been subtracted at the flatfield level for images taken in the $Y, J_1$, and $J_2$ filter where the stray light,
if not corrected, would have contributed an additive 5\%, 5\%, and 10\% to the flux, respectively. No stray light correction has
been done for the $H$-band images where the additive contribution is negligible at $<0.5\%$.    

From the coadded $H$-band images of the three pointings we created an $H$-band mosaic image with a total exposure time of 11600 sec pixel$^{-1}$ and 
1.03$\arcsec$ seeing on average. The summation process assigned a weight to each input image according to its transmission, background noise and PSF, 
so that the $H$-band mosaic has an optimal PSF. Using SExtractor \citep{Bertin} with default parameters we obtained a deep $H$-band source catalogue 
with 31747 objects. The astrometry was performed with IRAF \citep{tody93} using hundreds of bright ($H\lesssim16\fm 0$) 2MASS point sources in common 
with our catalogue reaching an accuracy of 0.1$\arcsec$ in RA and DEC.

The optical photometry has been re-derived for the source positions in the $H$-band catalogue instead of the previous $R$-band catalogue. Hence, optical and NIR
photometry in all 21 bands of the \C~ survey are measured in apertures matched in location and in size. In practice, we use a Gaussian weighting function in
our aperture to give more weight to the bright central parts of an object and less weight to the fainter outer parts. Using a Gaussian sampling function and
assuming a Gaussian seeing PSF allows us to sample identical areas of an object independent of seeing: we adjust the width of the Gaussian aperture to counteract
seeing changes such that the integral over the aperture is conserved under seeing changes. Mathematically, our brightness measurements are identical to placing
apertures with a Gaussian weighting function of $1\farcs7$ FWHM onto a seeing-free image in all bands (COMBO-17 chose $1\farcs5$ given the slightly better seeing
of its optical data). The NIR filters $Y$, $J_1$, $J_2$ and $H$ reach 10-$\sigma$ aperture magnitude limits of $22\fm 1$, $21\fm 5$, $21\fm 4$, and $21\fm 0$, respectively.

The COMBO-17 spectrophotometric stars have been used to ensure the calibration of the optical bands between each other \citep{wolf01b}.
We extended the calibration into the NIR using the \cite{pickles98} spectral library by visually matching the colours of point sources in our data 
to main sequence stars in the library. We estimate that the relative calibration between optical and NIR has a limited accuracy on the order of 7\%.
The calibration of our photometry in the $H$-, $J_2$-, and $J_1$-band has been verified by comparing the $H$ magnitude and the average value of
the $J_2$ and $J_1$ magnitude respectively to the $H$ and $J$ magnitudes of
340 stars with $H\lesssim16$ in common with the 2MASS catalogue. 
For the point sources, we found a mean offset in magnitude of $<1\%$ in the $H$-band and $<0.4\%$ in the averaged $J_1$- and $J_2$-band.
No verification could be done for the $Y$-band magnitude since no point source catalogue currently exists in this waveband.

\section{METHODS}

\subsection{Photometric Classification and Redshifts}

Photometric redshifts were determined using the multi-colour classification code by \cite{wolf98} as it was done in COMBO-17. 
Objects are divided into the four classes star, white dwarf, galaxy and quasar by comparing measured colours with colour libraries calculated from spectral templates.
The templates for stars and quasars are identical to those in \cite{wolf04}, while a modified library has been built for galaxies, in order to allow reliable estimates of
the stellar mass (see below). At the bright end, the redshift accuracy is limited by systematic errors in the relative calibration of the different wavebands or in a 
mismatch between templates and observed spectra. At the faint end, photon noise dominates. 

This dataset is a superset of the COMBO-17 data with the four NIR bands added. As a result, the photo-z's have changed very little at $z<1$, where
the NIR bands add little constraints, but it is reasonable to assume that they have improved over the optical-only results at $z>1$. In Fig.~\ref{new4SEDs.eps}
we show four examples of red galaxy SEDs and their best-fit templates. At $z<1$ the fit is clearly constrained by the original COMBO-17 data at $\lambda <1\mu$m.
At $z>1$ the four NIR bands act in concert with the optical data, while for red galaxies towards $z=2$ they are the sole providers of significant flux detections.
The two $z>1$-galaxies shown are examples of EROs with $R-H\sim 5$ and $5.7$, respectively. To the eye, their redshift is clearly constrained by locating the break
between neighbouring pairs of filters, while the fit takes all filters into account to constrain the redshift further.

COMBO-17 photo-z's have been shown to be accurate to $\sigma_z/(1+z)<0.01$ at $R<21$, $<0.02$ at $R<23$ \citep{wolf04}, albeit on a different field (the
CDFS). On the A901 and S11 field, we only have spectra for galaxies at $z<0.3$. From these, the photo-z dispersion of the cluster A901/2 has been
measured as 0.005 rms \citep{wmr05,gray09}. Presently, we lack the ability to confirm our photo-z's at $z>1$ here.
Hence, we need to rely on the plausibility of the SED fits to the photometry as shown in Fig.~\ref{new4SEDs.eps}.

However, the photometric errors and the grid of templates allow the estimation of probability distributions and confidence intervals in redshift for each galaxy.
These estimated redshift errors are shown in Fig.~\ref{newzerrors}, where they illustrate the change in behaviour with redshift and magnitude. Obviously, photometric
errors increase towards faint magnitudes and propagate into redshift errors, but at $z<1.2$ the redshift error is still very much driven by the deep optical photometry.
Hence, the bulk of objects has $\sigma_z/(1+z)<0.05$ even at our adopted $H$-band limit. At $z>1.2$, however, where the main redshift constraints are in the NIR bands,
a clear upswing of redshift errors with $H$-band magnitude can be seen.

\subsection{Galaxy Samples}

After eliminating $\sim 2300$ objects classified as non-galaxies, we defined our sample by first applying an $H$-band selection 
above our 5-$\sigma$ detection limit of $H=21\fm 7$. However, we are concerned with the evolution of galaxy samples that need to be complete in
rest-frame $V$-band luminosity or stellar masses. We avoid a colour bias by applying further magnitude cuts and eliminate from our colour-magnitude diagrams
those faint tails of the galaxy distribution that are known to be incomplete. Across our redshift range several observed bands with different completeness
limits map onto the rest-frame $V$-band. Hence, we opted to apply the following two further cuts: 

(1) At $z<0.43$ the entire observed $R$-band is redwards of the 4000$\AA$-break and roughly coincides with the rest-frame $V$-band. Here,
we require galaxies to have $R<23\fm 5$, which is the completeness limit of the optical-only COMBO-17 redshifts for $z<0.43$-galaxies. The presence of NIR
data may have deepened our completeness in this regime, but we opt to err on the conservative side and eliminate this faint end from our analysis. Note,
that optically-faint red galaxies at higher redshift (such as EROs) are by definition NIR-bright and have thus very well-constrained SED fits and (higher) redshift estimates
(see Fig.~\ref{new4SEDs.eps}).

(2) In the redshift range $0.43<z<1.4$, the $Y$-band is redwards of the break and we require $Y<22\fm 8$ (our 5-$\sigma$-limit). Again, objects that are particularly red in
$Y-H$ are expected to reside at $z>1.4$.

Finally, at $z>1.4$ the SED around the 4000\AA -break is entirely sampled by our NIR filters and the $H<21\fm 7$ selection is sufficient to ensure completeness and
an accurate redshift estimate. These selections are complete in stellar mass $\log (M_*/M_\sun)$ to 8.5, 9.5 and 10 in the redshift ranges $z<0.43$, $0.43<z<1.4$, and $z>1.4$, respectively. 

After removing galaxies with bad flags, our sample contains 10692 galaxies. Fig.~\ref{newzH} shows this sample and reveals already a number of large-scale structures. The
dominant overdensity at $z\sim 0.16$ is the original main target of this COMBO-17 field, the supercluster region A901/2 with well-over 1000 galaxies. Further clusters and
large-scale structures have been identified and reported in this field, partly from weak gravitational lensing \citep{taylor04,simon10} and partly from a galaxy cluster
search \citep{falter10}. These include localised clusters embedded in large-scale structures at $z\sim {0.26,0.37,0.5,0.7,0.8}$. At higher redshift and fainter magnitudes,
i.e. $z>0.8$ and $H>20$, redshift focussing effects are possible, and structures which appear only at faint magnitudes but without a sharp tail to the bright edge are unlikely
to be physically localised overdensities. In Fig.~\ref{newzH} the two possible unreal overdensities are the faint-end blobs at $z\sim 1$ and $z\sim 1.2$. Their appearance
does not require redshift errors in excess of what is discussed above, but only a mild focussing within the allowed errors. At $z>1.2$ no structure can be seen, as any physical
contrast has been smoothed by our redshift errors.

The histogram of photometric redshifts of the complete galaxy sample is shown in Fig. \ref{f1.eps}. Here, we show first the optical-only COMBO-17 redshifts of four different
fields, which demonstrate the signature of abundant large-scale structure. In case of the CDFS, these have been reported consistently by a variety of groups. Overall, the
structures include six Abell clusters as well as further clusters and rich groups, usually characterised by a pronounced red sequence. It is thus clear that we investigate
a variety of environments across redshift, and only the combination of several fields will eventually suppress error propagation from field-to-field variation.

The histogram in the right panel shows new redshifts after including the NIR bands of COMBO-17+4. It is selected by combined apparent-magnitude cuts in the $R$-, $Y$- and $H$-bands,
instead of the sole $R$-band in the optical-only photo-z's, and thus numbers at higher redshifts are increased. The inclusion of NIR data allows for good photo-z's at $z>1$,
but presently such data are only available for three quarters of the A901-field covering 0.2$\sq\degr$ area. This is our first look at $z>1.2$-galaxies, while completion of
our project will entail three fields with a different mix of environment at each redshift.
 
The bulk of our sample is in the range $0.7<z<1.1$, consistent with the $z=0.9$-peak in the $n(z)$ found in COSMOS \citep{ilbert10}.

\subsection{Spectral Library of Galaxies}

The photometric redshifts of galaxies in the optical-only \CB\ \citep{wolf04} are based on synthetic templates produced
by the population synthesis model PEGASE \citep{fioc97}. A single, exponentially declining burst of star formation was assumed for the star
formation history ($\tau=1$\,Gyr). The templates form a 2-dimensional grid in which one parameter is the age since the start of the burst
(60 steps between 0.05 to 15 Gyr) and the second one is extinction, which was modelled as a foreground screen of dust following the Small
Magellanic Cloud (SMC) law defined by \cite{Pei92}  in six steps from $A_V=0$ to $A_V=1.5$. Here, we employ a new library based on a
two burst-model following the approach of \cite{borch06}. The motivation for a two-burst model is to reproduce better the 
stellar population of blue galaxies that contain both an old stellar population and ongoing star-formation activity and to
provide more realistic mass-to-light ratios. However, the two-burst model by \cite{borch06} failed to deliver accurate redshifts. 

Our new two-burst library solves this problem by including extinction again and adjusting the age and strength of the second burst. 
For red galaxies (ages $> 3$\,Gyr) we leave the single-burst templates unchanged ($\tau_1=1$\,Gyr), and for bluer galaxies we add
a recent burst ($\tau_2=0.2$\,Gyr) to the old population starting 2.75\,Gyr after the first one. 
Increasingly blue galaxy templates are generated by both increasing the relative strength of the second
burst {\it and} by moving the final age (at which the galaxy is observed) from 3.0\,Gyr to 2.80\,Gyr ({\it i.e.} towards the start of
the second burst). This ensures that the templates fall into a region of the rest-frame colour-colour diagrams that is occupied by 
observed galaxies, and thus leads to very accurate photometric redshifts. We assume a Kroupa IMF, an initial metallicity 
of 0.01 ($\sim 2/3\, Z_\sun$) and neither infall nor outflow. 
More details will be given in Meisenheimer et al. (in preparation).

\subsection{Stellar Mass Estimation}\label{mass}           

The stellar mass of a galaxy is derived from the best-fitting template SED, whose characteristics are constrained mostly by the rest-frame
spectrum between 280~nm and the $V$-band, which is observed at all redshifts. Formally, we use data from all bands, but our masses can be 
seen to be derived from a galaxy's observed $V$-band luminosity and the $V$-band mass-to-light ratio of its best-fitting template.

Our use of a fixed rest-frame band to determine a galaxy's luminosity differs from that of \cite{borch06} who only used the reddest observed-frame 
band with good photometry, and thus may suffer unintended bias trends with redshift. At low redshifts $z<0.5$, \cite{borch06} sample the 
rest-frame $V$-band as well so that our mass estimate should only differ from theirs due to the different templates. Only due to the NIR
extension in \C~ can we continue to use the rest-frame $V$-band out to $z=2$. 

Indeed, a comparison of galaxies at the redshift of the super-cluster Abell 901/2 ($0.150<z<0.175$) shows that systematic differences 
in the stellar masses do not exceed 0.1\,dex. Beyond $z=0.5$ the stellar masses derived here are superior due to the NIR photometry. We estimate 
a typical mass accuracy on the order of 30\% across all redshifts.

We wish to assess possible biases in the stellar mass estimates arising from dust and consider the \citet{belldejong01} equation, 

\begin{equation}
\log_{10} M_*/M_\odot \propto -0.4 M_V+1.305 (B-V)
\end{equation}

because it allows an analytic derivation of a reddening vector in colour-mass diagrams, and our masses are consistent with Eq.~1 within $\pm 0.1$~dex. Any dust reddening
can now be seen as a mean reddening plus some structure. Approximating the mean reddening by a uniform screen of dust right in front of the stellar population changes
its colour by $\Delta(B-V)=E_{B-V}$ and its $V$-band luminosity by $\Delta V=R_V E_{B-V}$. The reddening-induced overestimate of the $M/L$-ratio and the absorption-induced
underestimate of $L$ cancel exactly, if

\begin{equation}
0.4 R_V E_{B-V} = 1.305 E_{B-V} \rightarrow R_V = 3.2625
\end{equation}

which is almost true for the Milky Way, LMC or SMC dust law from \citet{Pei92}. Dust acting like a foreground screen on the stellar population will thus not bias the mass
estimates at all. However, if pockets of highly absorbed stars exist as well, they will be entirely withdrawn from the optical view and not contribute to either luminosity or
colour. They will thus be plainly not taken into account and the final mass estimate will be underestimated by just the stellar mass present in highly obscured pockets.

\subsection{Rest-Frame Luminosities and Colours} \label{restframe}

Rest-frame luminosities are derived from the observed photometry covering the wavelength range from the $U$-band to the $H$-band. 
We obtained rest-frame luminosities for the $V$-band in the Johnson photometric system as well as for the $U_{280}$-band, a synthetic UV 
continuum band with a top-hat transmission curve that is centred at $\lambda =280$\,nm and $40$\,nm wide. These two filters are covered by our observed bands across almost
the entire range of interest. This is essential as our study does not rely on SED extrapolation except for galaxies in the small range $0.2<z<0.3$. The rest-frame colour
$U_{280}-V$ is measured robustly against redshift errors since both filters are located in smooth continuum regions of the galaxy spectra. The Johnson $U$-band,
in contrast, partly overlaps with the 4000\AA -break and is hence affected strongly by small uncertainties in the redshift determination.

The rest-frame luminosities are derived as described by \cite{wolf04}. The best-fitting SED is placed into the aperture photometry and
integrated over the redshifted rest-frame bands. We have taken into account the interstellar foreground extinction as well as a correction 
from aperture to total photometry, which could be biased by colour gradients. It is determined from the total magnitude MAG-BEST derived by 
SExtractor on the $H$-band image and thus certainly correct for any rest-frame band overlapping with the observed $H$-band. The magnitude errors are determined from
a propagation of photometric errors. They include a minimum error of $0\fm 1$ to take into account redshift errors and overall calibration uncertainties, for details
see \cite{wolf04}.

\subsection{Error-Weighted Colour Histograms}\label{methodwhisto}

Generally, colour and magnitude information can be used to separate passive red galaxies from star forming blue ones. However, at high redshift the
well-known colour bimodality may be blurred by scatter from relatively large errors on the colour measurements. We try to manage this challenge and
investigate the galaxy colour bimodality through cosmic time by using error-weighted colour histograms. This method represents each galaxy by its 
Gaussian probability distribution in colour $c$, i.e. $p(c)\sim e^{- (c-c_0)^2/2\sigma _c^2}$, where $\sigma _c$ is the colour error and $p$ is normalized so that $\int p dc$=1. 
Thus, a galaxy with a small error has a more peaked distribution and contributes more structure to the summed distribution than a galaxy with a large error. 
As a result, the structure in our histograms is driven entirely by galaxies with small colour errors, while objects with large errors lift the overall counts without
producing peaks. We produced error-weighted colour histograms by summing all the Gaussian distributions within many thin redshift slices ($\Delta z\sim0.1$) stepping
through our full redshift range of $0<z<2$. Towards highest redshifts, the increasing colour errors will dilute the contrast of the red-sequence peak. Redshift errors
can also lead to some spill-over from a physical redshift bin into neighbouring photo-z bins and produce scatter in our measured colour evolution from bin to bin in redshift.

To better disentangle the red sequence from the blue cloud we have tilted the $(U_{280}-V)$ colour in the colour-magnitude plane, see Eq. \ref{Eq.1}. 
The measured $(U_{280}-V)$ of each individual galaxy is projected along the slope of the red sequence as determined in the CMD of the super-cluster 
A901/2 (see Fig. \ref{f5.eps}, top left) to the pivotal magnitude $M_V=-20$: 

\small
\begin{equation}
(U_{280}-V)_{M_V=-20}=(U_{280}-V) + 0.3(M_V +20)
\label{Eq.1}
\end{equation}
\normalsize

The derived slope of 0.3 is consistent with that of 0.08 in the $(U-V)$ colour in \cite{bell04}.

\section{EVOLUTION OF THE COLOUR BIMODALITY}

\subsection{Rest-Frame Colours and Colour Errors as a Function of Redshift}\label{colourerror}

Scatter in measured galaxy colours affects the appearance of the colour bimodality in a colour-magnitude diagram and may render it invisible,
especially in a sample of low density field galaxies where no clusters rich in red galaxies produce a well-defined sequence. As the scatter in
colour increases with redshift, this effect can prevent us from detecting a physically present high-redshift bimodality.

For a first assessment of our ability to trace bimodality, 
we plot the rest-frame $(U_{280}-V)$ colour and its error $\delta(U_{280}-V)$ as a function of redshift in Fig. \ref{f2.eps}. 
In panel (a), we see the number of red galaxies decreasing with redshift. Also, a lack of star-forming blue galaxies with $(U_{280}-V)<1$ 
at $z>1.5$ is caused by a red-object bias from our $H$-band selected catalogue. Panel (b) shows that the colour accuracy is very good for 
objects located at low redshift $z\lesssim0.9$. The bulk of the galaxies have a colour error close to the assumed minimum of $0\fm 1$. 
The large scatter around $z\sim 1$ is caused by a locally increased magnitude error in the rest-frame $V$-band that results from $M_V$ being 
calculated from the relatively shallow narrow-band $J_1$. In contrast, the
$U_{280}$ filter overlaps with the $\sim 3$ mag deeper $R$-band at $z\sim 1$. 
At $z>1.7$, the colour uncertainties grow larger again since the rest-frame where the $U_{280}$-band falls into the $I$-band, which is our shallowest broad-band.

\subsection{Colour Bimodality and the Emergence of the Red Sequence}\label{whisto}

In Figure \ref{f3.eps} we plot error-weighted colour histograms of galaxies in different redshift slices. Each panel shows two distinct peaks due to the bimodality,
whereby the right peak represents red-sequence galaxies sitting on top of a tail of blue cloud galaxies reaching smoothly towards the red due to lower star-formation rates
or dust reddening. Clearly, a red galaxy sample defined by a colour cut will comprise both quiescent red-sequence galaxies as well as dusty star-forming galaxies. At high
redshift, we are unable to separate old red from dusty red galaxies, and so a red galaxy sample overestimates the space density of quiescent galaxies. However, the
colour-magnitude relation should be well constrained, as only galaxies from the proper quiescent red sequence are focussed in colour and contribute to the peak in the
colour histogram, while dusty red galaxies form a smooth underlying continuum spreading across the red-sequence and extending beyond.

The super-cluster A901/2 seen in the top left panel has a particularly clear red sequence. 
Our results show a clear galaxy bimodality at $z<1$ as already established by \cite{bell04}. However, the depth of our survey allows us to extend the 
detection of the galaxy bimodality up to $z\sim1.65$, the mean redshift of the highest interval where we can still detect two distinct distribution peaks. 
Beyond $z\sim1.65$ it is not possible to confirm a galaxy colour bimodality since the number of objects available for our analysis drops considerably. 
Figure \ref{f3.eps} bottom right shows the distribution of 423 galaxies in the redshift interval $1.7<z<2$. It is clearly not bimodal, though this does not mean
that the red sequence does not exist there. Our redshift errors are unlikely to be the main source of this disappearance, since the redshift error $\sigma_z/(1+z)$ of
luminous red galaxies with $M_V=[-23.5,-22]$ degrades only from a median value of $\sim 0.01$ at $z=0.8$ over $\sim 0.025$ at $z=1.2$ to $\sim 0.04$ at $z=1.6$.
{Instead, it is the combination of a relatively small number of quiescent luminous red objects due to the small area surveyed and the washing out of any red sequence signal due
to increasing colour errors (see Fig. \ref{f2.eps}b) that prevents us from detecting the red sequence beyond this redshift.

\subsection{Evolution of the Colour-Magnitude Relation}\label{zcmr}

We use the error-weighted colour histograms in thin redshift slices of $\Delta z\sim0.1$ to track the colour evolution of the peak in the red sequence
and plot $(U_{280}-V)$ at the pivot point $M_V=-20$ in Figure \ref{f4.eps}. The colour of the red-sequence peak is obtained with the MIDAS
routine \texttt {CENTER/GAUSS gcursor}, which fits a Gaussian function to an emission line on top of a continuum. We mark the 'continuum' level on both sides of the peak
interactively, and obtain the peak location from the fit. We obtain errors from the uncertainty of the peak position obtained by varying the interval in which the Gaussian
is being fit, but we do not include systematic calibration uncertainties in colour (see Table \ref{tbl-2} for results). We find that the $(U_{280}-V)_{M_V=-20}$ colour of 
the peak evolves almost linearly with lookback time $\tau$, and obtain a linear best fit (dashed line) of
\begin{equation}
(U_{280}-V)_{M_V=-20}=2.57-0.195 \times \tau
\label{Eq.2}
\end{equation} 

Our data points of the peak colour scatter somewhat around this fit,
  which is an indication of our uncertainties in measuring the colour, as any
  cosmological evolution of the average galaxy population is expected to be
  monotonic. However, these variations may also result from us looking at
  different environments at different redshift, which may be at different
  evolutionary stages, and not solely from our methodical uncertainties.

We obtain a colour-magnitude relation (CMR) as a function of redshift using the approximation for the lookback time $\tau \simeq 15 Gyr \cdot z\cdot(1+z)^{-1}$, and separate 
blue and red galaxies with a parallel relation 0.47 mag bluer than the CMR, which is
\small
\begin{equation}
(U_{280}-V)_{lim}=2.10-0.3(M_V+20)-2.92 z/(1+z)
\label{Eq.3}
\end{equation}
\normalsize 

In the low redshift regime $0<z<1$ this cut is consistent with the one derived by \cite{bell04} considering an approximate colour transformation 
derived from our templates, which is $(U-V)=0.28+0.43(U_{280}-V)$. Our CMR is slightly steeper than the one derived by \cite{bell04}, but that 
translates only into a minor colour difference of $<0.1$ mag for galaxies in the relevant magnitude range $-24<M_V<-18$. Although we see no clear red 
sequence at $z\gtrsim1.65$, we extrapolate the CMR all the way up to $z=2$ to isolate red galaxies across our entire sample. 

The solid line in Figure \ref{f4.eps} shows the evolution predicted by PEGASE for a single stellar population formed at a lookback time 
of 12 Gyr ($z_f$ = 3.7) with solar metallicity. The bulk of our data are
consistent with the pure aging model. Our lowest-redshift value is based
entirely on the A901/2 super-cluster at $z=0.165$ (Lookback time=2.0 Gyr),
which is the highest-density environment in our sample.

\section{EVOLUTION OF THE RED-SEQUENCE GALAXY POPULATION}\label{EvolRS}

We now focus on the colour and the mass evolution of the whole red galaxy population, and analyse colour-magnitude diagrams (CMD) and 
colour-stellar-mass diagrams (CM$_*$D) in redshift slices. We divide our galaxy sample into red and blue populations with the cut in Eq. \ref{Eq.3}. 
Across the entire sample at $0<z<2$, roughly a third of the galaxies are red (3163 out of 10692), but in the high-redshift part at $z>1$ less
than a quarter are red (843 out of 3479 galaxies).

\subsection{Colour-Magnitude Diagrams}

The evolution of the red-sequence population is presented in the CMDs of Fig. \ref{f5.eps}. In each panel, the solid line indicates the 
bimodality separation of Eq. \ref{Eq.3} assuming the mean redshift of the slice. The colour of the data points shows their individual 
nature, and due to the width of the redshift intervals some galaxies scatter across the separating line.
The top left panel in Fig. \ref{f5.eps} shows the CMD of the A901/2 super-cluster centred at $z=0.165$. Such a dense environment
shows a clear red sequence, and this was used to derive the red sequence slope in Eq. \ref{Eq.1}. The red sequence is not as sharp
and easy to visually distinguish on CMDs beyond $z=1$, especially in the redshift slice $1.07<z<1.19$ due to the local increase in the
scatter caused by large colour uncertainties. This highlights the necessity of using error-weighted histograms to derive the CMR.

Altogether, we find that at a given magnitude both galaxy populations were bluer in the past, and in particular the bright end of the red 
sequence became $\sim0.4$ mag redder from $z=2$ to $z=0.2$. We also find an increasing population of bright ($M_V<-22$) blue ($(U_{280}-V)<1$) 
galaxies at $z>1$, which has also been reported by \cite{taylor09} in $(U-R)$-vs.-$M_R$ CMDs of a $K$-selected sample at $0.2<z<1.8$ in 
the Extended Chandra Deep Field South.

\subsection{Colour-Stellar Mass Diagrams}

Fig. \ref{f6.eps} shows the evolution of the population in CM$_*$D for the same redshift slices as the CMDs in Fig. \ref{f5.eps}. 
We select the red sequence (black points) again with Eq. \ref{Eq.3}. We reach smaller masses than \cite{borch06} because our galaxy sample 
is primarily $H$-band selected instead of $R$-band selected. Conversely, our blue galaxy population reaches less deep than that of \cite{borch06}. 
We see a general trend whereby at fixed mass both galaxy populations were bluer in the past, just as they were at fixed magnitude. 
We find that massive galaxies between $0.2\lesssim z \lesssim 1.0$ are dominated by the red population. This was also observed by
\cite{borch06}, who derived mass functions for the red and the blue galaxy
population at $z<1$ using the optical \CB \ data, and
by many other authors \citep[see e.g.][]{bundy05}. However, at $z>1$ there is a growing
population of massive ($\log M_*/M_\sun \gtrsim 11$) blue ($(U_{280}-V)\lesssim 1$) galaxies as we go back in time, a phenomenon
\cite[also observed by][]{williams09,taylor09} that has no analogue in the local Universe.

The bottom right panel shows our highest redshift slice ($1.78<z<2$) where our sample contains 45 red galaxies. Among them are eight very massive 
($\log M_*/M_\sun >11.5$) objects, which means that at $z\sim2$ the red sequence already contains very massive galaxies. The most massive
object found in this high redshift slice has a stellar mass of $\log M_*/M_\sun =12.0$. Our red sequence will contain dusty star-forming galaxies
besides old galaxies, but since these do not form a sequence and are not focussed in colour, they will not affect the measurement of the red-sequence
colour with our tailored method.

\subsection{Number Density Evolution of Massive Red Galaxies}

Figure \ref{f7.eps} shows the evolution of the number density of massive red
galaxies as a function of redshift (see also Table \ref{tbl-3}). These red
galaxies include both quiescent old galaxies as well as dust-reddened galaxies.
We avoid the super-cluster A901/2 by constraining our sample to $0.2<z<2$ and choose masses of $\log M_*/M_\sun >11$, where we are complete 
at all redshifts, thus retaining only 478 objects.

We find that the number density of massive red galaxies rises considerably (by a factor $\sim4$) from $z\sim 2$ to $z\sim 1$ and is more or
less constant at $z<1$. Due to the small survey area our results are affected by cosmic variance on the order of 30\% as estimated using 
\cite{moster10}. Nevertheless, our results are comparable within the error bars to results by the GOODS-MUSIC survey 
(0.04 deg$^2$ area) \citep{fontana06} and the MUSYC survey (0.6 deg$^2$ area) \citep{taylor09}. In the highest redshift bin at $z\sim2$
the number density derived is consistent with the spectroscopic survey of \cite{kriek08}. Since any sample of red galaxies could in principle
be contaminated by some dusty red galaxies the number density derived here represents an upper limit. However, we do not expect the effect to 
be large at the high-mass end.

\subsection{Stellar Mass Density Evolution of Massive Galaxies}

The different evolution of stellar mass density in red and blue galaxies is shown in Fig.~\ref{f8.eps} 
(see also Table \ref{tbl-3}). Again we restrict ourselves to masses of $\log M_*/M_\sun >11$, where we are complete across $0.2<z<2$.
Like the number density, the stellar mass density of the red population increases by a factor of $\sim4$ from $z\sim 2$ to $z\sim 1$ 
and remains roughly constant at lower redshifts. In contrast, the stellar mass density of the blue population increases only by a factor of 
$\sim2$ from $z\sim 2$ to $z\sim 1.2$ and decreases by the same factor towards low redshift again. Altogether, the overall stellar mass 
density increases by a factor of $\sim$3 from $z\sim 2$ to $z\sim 1$ due to the combined contribution of red and blue galaxies, while it 
remains constant at lower redshift where the red sequence dominates. Thus, the main formation epoch of the massive red galaxy population 
is ranges over $2>z>1$.\\

Comparing our stellar mass densities with the literature is not foolproof since different surveys have different selection criteria 
using e.g. colour, morphological types and star formation activity. They employ different methods to derive the stellar mass and they
can be affected by cosmic variance.

Our results for the evolution of the mass density of the entire population between $1>z>0$ are consistent with \cite{conselice07} 
who also found little evolution for galaxies with $11<\log M_*/M_\sun<11.5$. At higher redshift, our results differ: \cite{conselice07} 
found an increases by a factor 10.7 from $z\sim 2$ to $z\sim 1$, while our results show a more modest rise by a factor $\sim3$.

Based on a $3.6\mu$-selected sample of galaxies \cite{arnouts07} found an increase of the mass density in the quiescent population 
by a factor of 2 from $z\sim 1.2$ to $z\sim 0$, while the star-forming population shows no evolution. Additionally, at higher redshift 
between $2>z>1.2$ they found that the quiescent population increases by a factor of 10 while the star formation population increases 
by a factor of 2.5. However, these results are based on a magnitude-selected sample and not a mass-selected one.

\cite{cirasuolo07} found in a sample of $M_k\leq-23$ galaxies that the space density of bright red galaxies is nearly constant over 
$1.5\gtrsim z \gtrsim0.5$, while that of bright blue galaxies decreases by a factor of $\sim2$ over the same redshift interval.

Recently, \cite{ilbert10} found a rise in the stellar mass density of $\log M_*/M_\sun >11$ galaxies from $z\sim 2$ to $z\sim 1$ 
by a factor $\sim 14$ for quiescent galaxies and a factor of $\sim4.3$ for red-sequence galaxies, similar to our result. 
This difference between quiescent and red-sequence galaxies is likely to arise from red star-forming galaxies that contaminate the 
red sequence more towards $z\sim 2$. For both quiescent and red-sequence galaxies, they found little evolution at $z<1$, as we do. 
Also, their highly star-forming sample compares well to our blue galaxy population. They find a rise in mass density by a factor of 
$\sim1.5$ from $z\sim 2$ to $z\sim 1$ and a decrease by a factor of $\sim2.5$ at lower redshift again.

\section{SUMMARY AND CONCLUSION}

In this work we investigated the evolution of the red sequence in terms of colour, luminosity, mass, and number density through cosmic time since $z=2$.
We derived an $H$-band catalogue of 10692 galaxies from 0.2 deg$^2$ of the A901-field surveyed by the deep NIR multi-wavelength survey \C. While deep multi-wavelength surveys
provide photometric redshifts for large samples of galaxies, the measured colours suffer from large uncertainties at high redshift that wash out the contrast with which the
red sequence appears on top of the tail extending from the blue cloud. Hence, we used colour histograms weighted by the colour error of each galaxy to trace even a diluted
red-sequence signal. We also used the rest-frame colour $(U_{280}-V)$ for maximum population contrast and minimum sensitivity to redshift errors.

As a result, we found a red sequence up to $z\sim1.65$, beyond which the situation is unclear. Tracking the colour evolution of the red sequence peak, we derived an evolving
colour-magnitude relation up to $z=2$ and used it to separate the red and blue galaxy populations. Our results show that the $(U_{280}-V)$ colour evolution of the red sequence
is consistent with pure aging. Further results are:

\begin{enumerate}
\item  Both the red and blue galaxy population get redder by $\Delta (U_{280}-V)\sim$0.4 mag since $z=2$.
\item  The population of massive blue galaxies grows from $z\sim 2$ to $z\sim 1$.
\item  The massive end of CM$_*$Ds is dominated by the red galaxy population at $z<1$ and by both galaxy populations at $z>1$.
\item  Some massive red galaxies with $\log M_*/M_\sun\sim 11.5$ are already in place at $z\sim2$.
\end{enumerate}

We investigated the number density and stellar mass density evolution of massive red galaxies and found that both increase by
a factor $\sim4$ between $2>z>1$ and shows little evolution since $z=1$. This suggests that the main formation epoch of massive red galaxies is at $2>z>1$ such that they have assembled most of their mass by $z\sim1$. Note, that our masses are unlikely to be biased much by dust.

It is clear that our results are affected by cosmic variance due to the small area surveyed. Once the data are available from the 
two other fields (A226 and S11) targeted by the \C~ survey, we expect to create an $H$-band selected catalogue of $\sim50000$ galaxies
in an area of 0.7 deg$^2$, of which $\sim12000$ galaxies will be above $z=1$. The more than threefold increase in area and the combination
of unrelated fields on the sky will reduce the effect of cosmic variance by a factor three and firm up our findings quantitatively.

\acknowledgments

We are grateful to all Calar Alto staff astronomers in particular Jesus 
Aceituno, Anna Guijaro, Felipe Hoyo and Ulli Thiele. Thanks to the International
Max-Planck Research School for fundings. CW was supported by an STFC Advanced Fellowship.

\clearpage



\begin{figure}
\begin{center}
\includegraphics[scale=0.5, angle=0, bb=132 148 538 763]{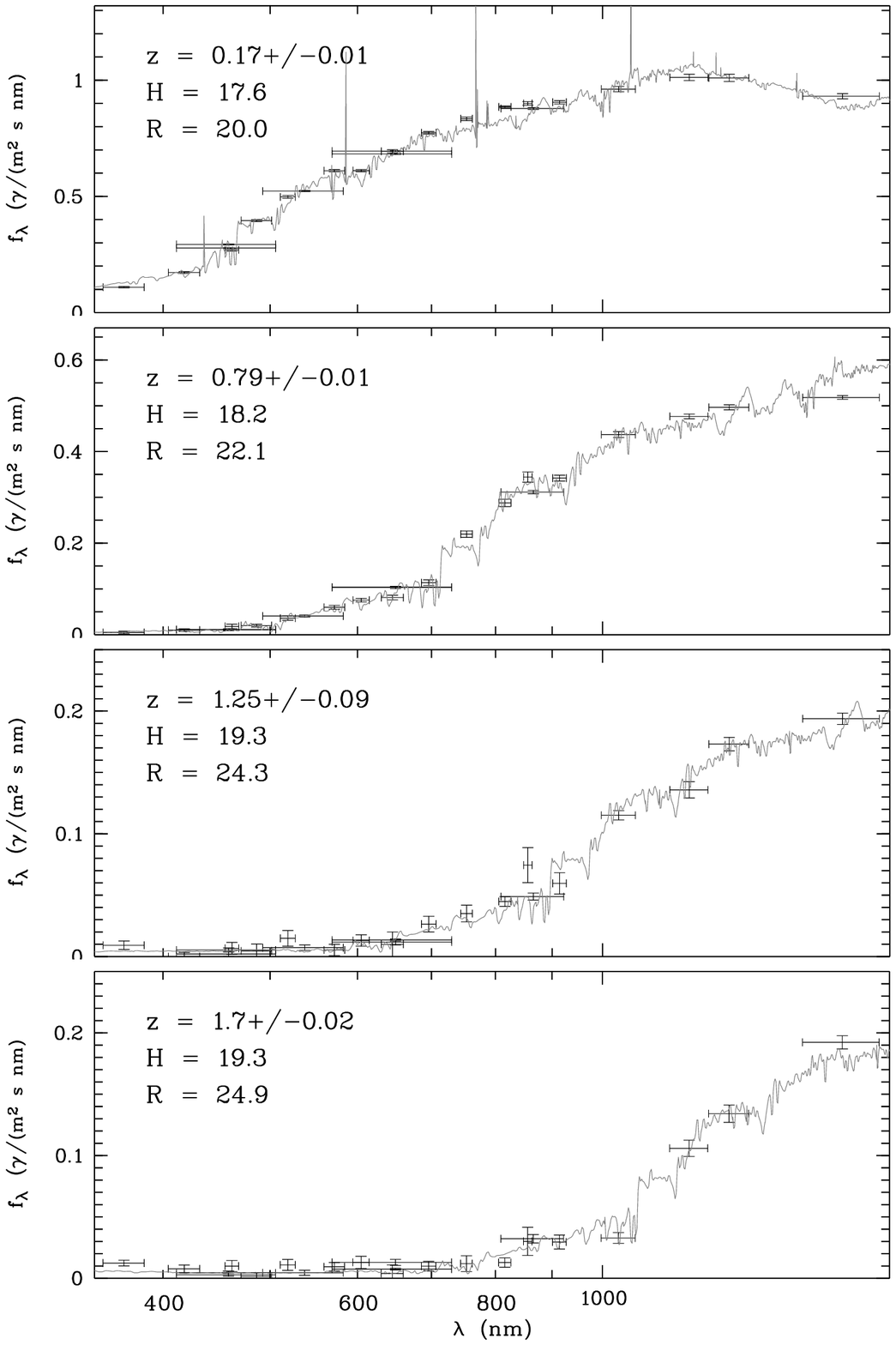}
\caption{\textit{Spectral energy distributions of four red-sequence galaxies at different redshift.\label{new4SEDs.eps}}}
\end{center}
\end{figure}

\begin{figure}
\begin{center}
\includegraphics[angle=270,clip,width=\hsize]{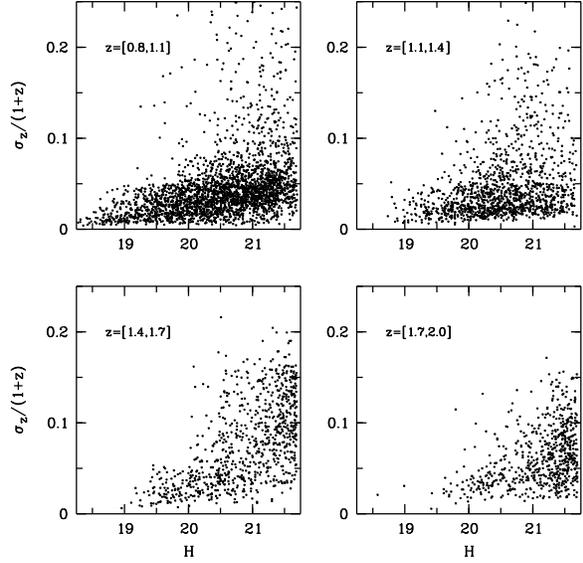}
\caption{\textit{Photometric redshift accuracy vs. H-band magnitude in higher-z redshift intervals.\label{newzerrors}}}
\end{center}
\end{figure}

\begin{figure}
\begin{center}
\includegraphics[angle=270,clip,width=\hsize]{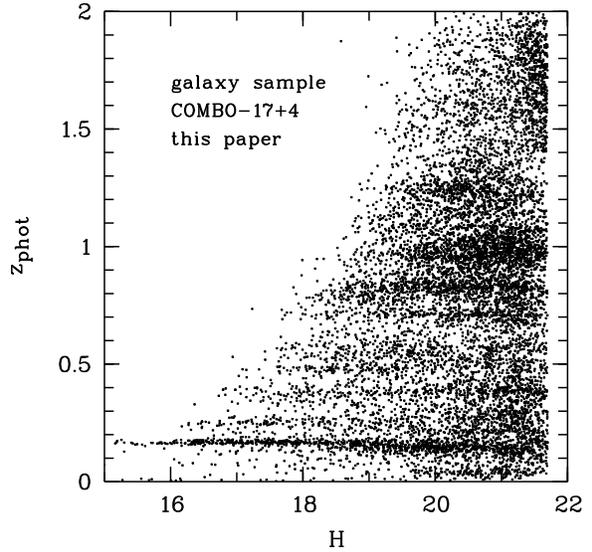}
\caption{\textit{Photometric redshift distribution of our galaxy sample vs. H-band magnitude.\label{newzH}}}
\end{center}
\end{figure}

\begin{figure}
\begin{center}
\includegraphics[angle=270,clip,width=\hsize]{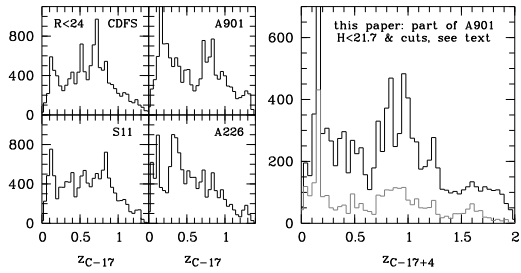}
\caption{\textit{Left: Original COMBO-17 photo-z distribution in four fields, illustrating field-to-field variation. Right: New COMBO-17+4 
distribution of the sample used in this paper (black line), and the red sample
as defined in Sect.~4.3 (grey line). \label{f1.eps}}}
\end{center}
\end{figure}

\begin{figure}
\epsscale{1}
\plotone{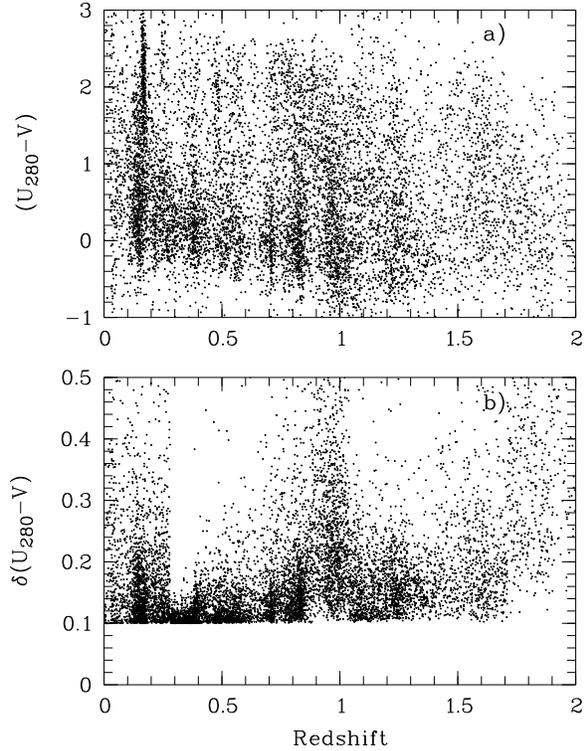}
\caption{\textit{Panel a): Rest-frame colour ($U_{280}-V$) over redshift. The A901/2 super-cluster produces a prominent feature at z=0.165. 
Panel b): Colour error over redshift, as determined from error propagation. At high redshift the colour determination is less accurate, especially at $z\gtrsim1.5$. An error floor of $0\fm 1$ is assumed, see Sect.~\ref{restframe}.\label{f2.eps}}}
\end{figure}


\begin{figure}
\epsscale{1}
\plotone{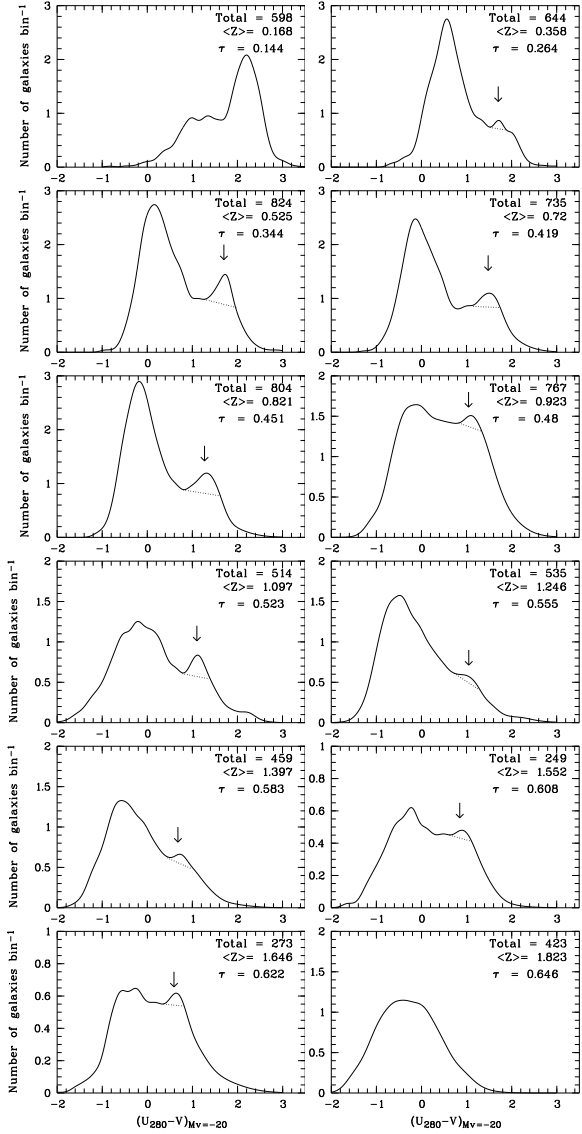}
\caption{\textit{Rest-frame $(U_{280}-V)_{M_V=-20}$ galaxies colour distribution in redshift slices. 
Histograms are sums of Gaussian probability distributions representing single galaxies with their individual colour error. 
Structure is driven by bright objects with accurate colours while faint, large-error objects with broad, smooth distributions don't add contrast
to the peaks. The top left panel contains the super-cluster A901/2 with its prominent red sequence.
The colour of the red sequence peak, indicated by an arrow, is determined by a Gaussian fit to the localised excess above a continuum of galaxy counts.
Thus, a red sequence is observed up to $z\simeq 1.6$. Number of galaxies, mean redshift and mean lookback time $\tau = z/(1+z)$ are noted.}}
\label{f3.eps}
\end{figure}


\begin{figure}
\begin{center}
\includegraphics[scale=0.5,clip=true,angle=0,bb= 0 0 488 373]{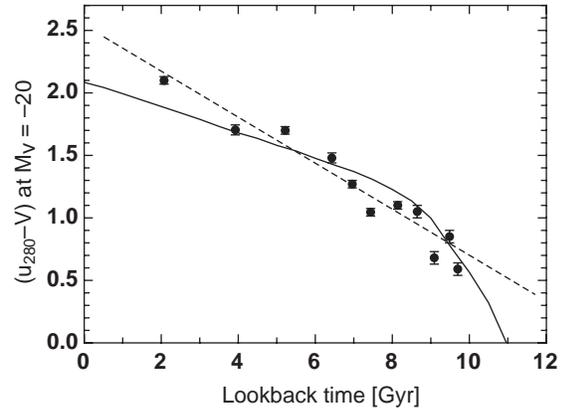} 
\caption{\textit{Evolution of the red-sequence colour with lookback time $\tau
    = 15~{\rm Gyr} \cdot z(1+z)^{-1}$, see Table \ref{tbl-2}. 
The dashed line is a linear fit to the points, and the solid line is an
example prediction using PEGASE for a single age stellar population with solar
metallicity formed 12 Gyr ago ($z_f=3.7$). The error bars account only for the colour measurement of the peak of the bright end of the red sequence. However, the scatter among the points illustrates further uncertainties including systematics and cosmic variance.}}
\label{f4.eps}
\end{center}
\end{figure}


\begin{figure}
\epsscale{1}
\plotone{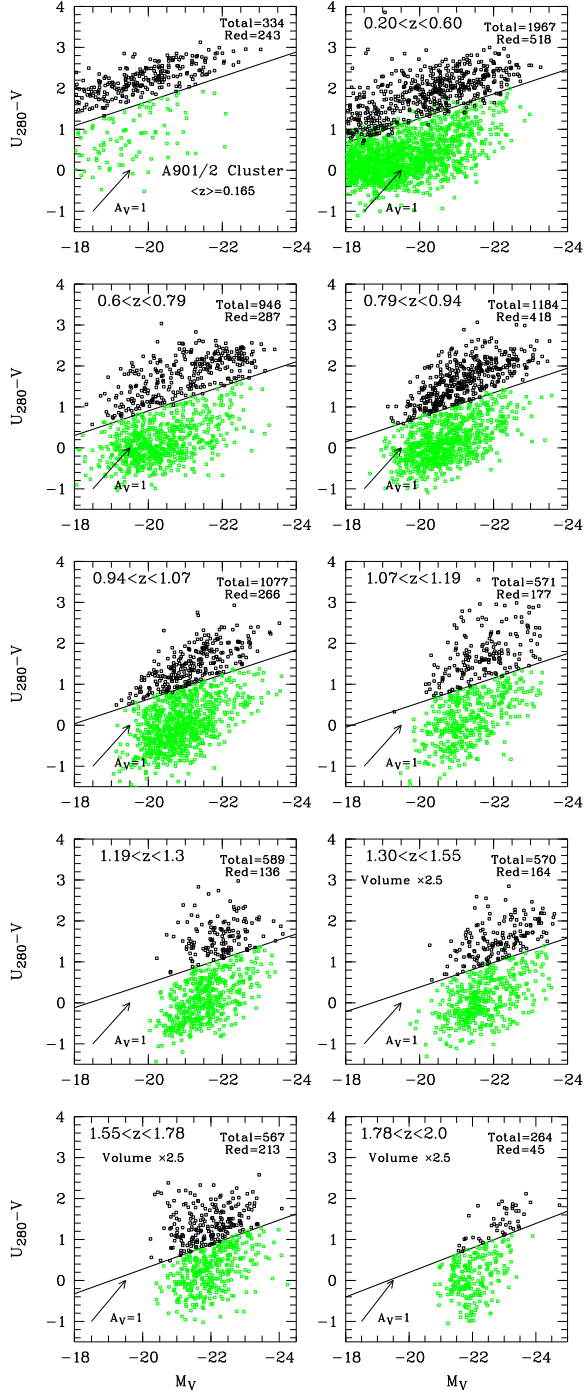}
\caption{\textit{Colour-magnitude diagrams in redshift slices. 
Solid lines indicate the bimodality separation (see Eq. \ref{Eq.3}) at the mean redshift of each bin. The real separation is given by the
colour of the data points (black=red sequence, green=blue cloud). Total number of galaxies and red fraction are noted. 
All redshift bins are nearly equal comoving volumes except for the three highest bins, which enclose 2.5 times as much volume, and the A901/2 cluster bin with a small volume enclosed within $0.150<z<0.175$. For reference, we show
a reddening vector for $A_V=1$ mag. (A colour version of this figure is available online.)}}
\label{f5.eps}
\end{figure}


\begin{figure}
\epsscale{1}
\plotone{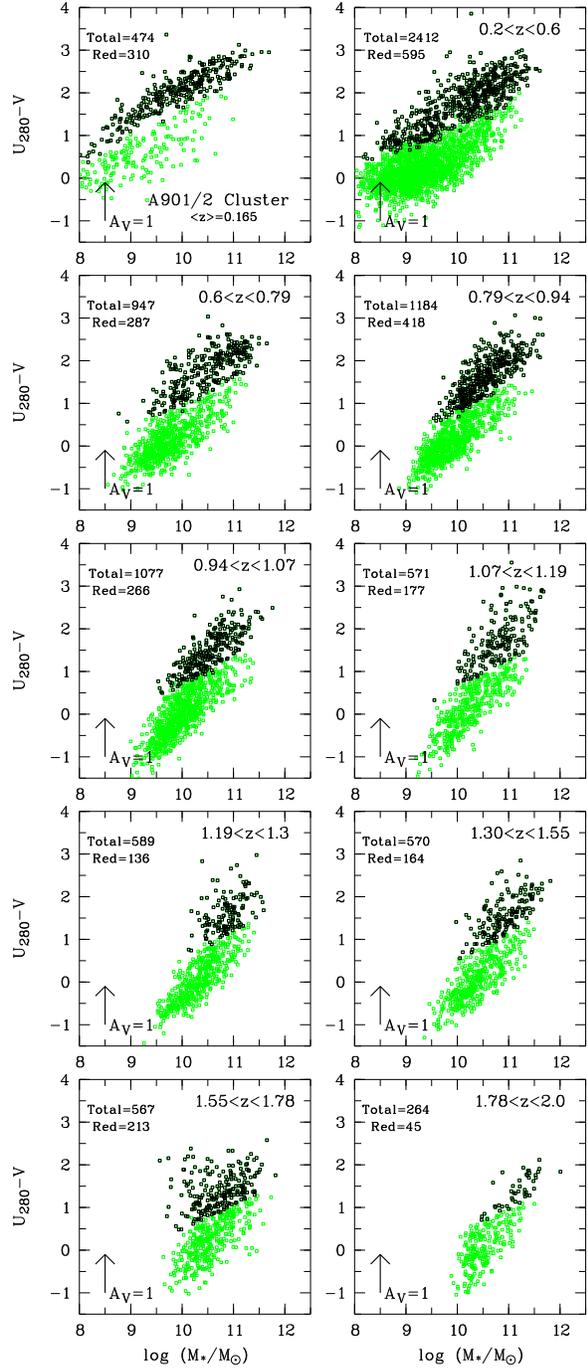}
\caption{\textit{Colour-mass diagrams in same redshift slices and format as Fig. \ref{f5.eps}. The reddening vector assumes a \citet{belldejong01} law for the stellar-mass-colour relation and a \citet{Pei92} dust law. (A colour version of this figure is available online.)}}
\label{f6.eps}
\end{figure}


\begin{figure}
\begin{center}
\includegraphics[scale=0.3,clip=true,angle=-90,bb= 49 80 566 775]{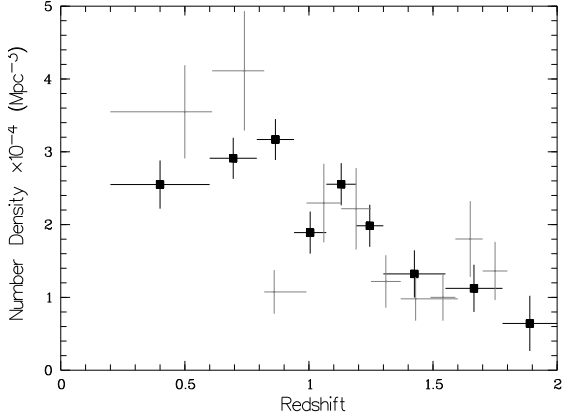}
\caption{\textit{Number density evolution of massive red galaxies with $\log M_*/M_\sun>11$ and $(U_{280}-V)>(U_{280}-V)_{lim}$ as a function
of redshift. The error bars represent the cosmic variance in each redshift bin and were calculated with a method described by \cite{moster10}.
In grey, data from \cite{taylor09} for comparison.}}
\label{f7.eps}
\end{center}
\end{figure}


\begin{figure}
\begin{center}
\includegraphics[scale=0.3,clip=true,angle=-90,bb= 49 29 566 775]{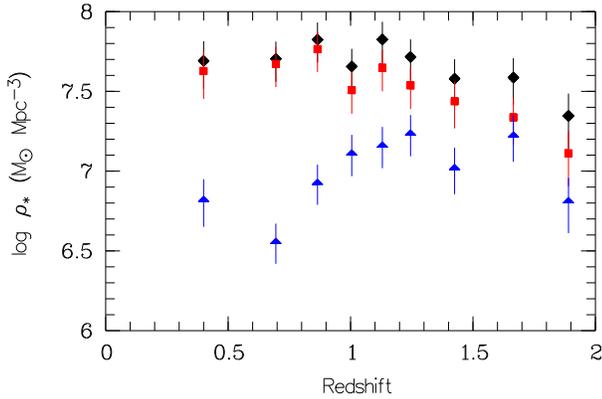}
\caption{\textit{Stellar mass density evolution of galaxies with $\log M_*/M_\sun>11$ as a function of redshift.
The entire galaxy population (diamond) is divided into the red-sequence (square) and blue-cloud galaxy population (triangle). 
Error bars include only cosmic variance estimates and no uncertainty from the stellar mass estimation. (A colour version of this figure is available online.)}}
\label{f8.eps}
\end{center}
\end{figure}

\clearpage



\begin{deluxetable}{ccccccc}
\tabletypesize{\scriptsize}
\tablecaption{COMBO-17+4 Target Coordinates and Integration Times\label{tbl-1}}
\tablewidth{0pt}
\tablehead{
\colhead{Field \tablenotemark{a}} & \colhead{RA} & \colhead{DEC} &\multicolumn{4}{c}{Integration Time per Filter\tablenotemark{b}}\\ 
               & \colhead{} &\colhead{} &\colhead{H} &\colhead{J$_2$} &\colhead{J$_1$} &\colhead{Y}\\
		& \colhead{(J2000)} &\colhead{(J2000)} &\colhead{(ksec)} &\colhead{(ksec)} &\colhead{(ksec)} &\colhead{(ksec)}                  
}
\startdata
\bf A901 &\bf  09h 56m 17s &\bf -10$\degr$ 01\arcmin 25\arcsec &\bf   11.6 &\bf 11.9 &\bf 8.6 &\bf 8.4\\
A226 &  01h 39m 00s & -10$\degr$ 11\arcmin 00\arcsec &   14.6 &17.3 &11.6 &8.8\\ 
S11  &  11h 42m 58s & -01$\degr$ 42\arcmin 50\arcsec &   16 &4.5 &8 &10\\
\enddata
\tablenotetext{a}{Results in this paper are derived from the observations of the A901-field only.}
\tablenotetext{b}{Integration time averaged over different subfields.}
\end{deluxetable}


\begin{deluxetable}{ccccc}
\tabletypesize{\scriptsize}
\tablecaption{$(U_{280}-V)_{M_V=-20}$ Colour of the Red-Sequence Peak\label{tbl-2}}
\tablewidth{0pt}
\tablehead{
\colhead{Mean Redshift} & \colhead{Redshift Interval} & \colhead{$(U_{280}-V)_{M_V=-20}$} & \colhead{$\delta$ $(U_{280}-V)_{M_V=-20}$} & \colhead{Lookback Time}\\ 
\colhead{}              & \colhead{}                  & \colhead{(Vega mag)}              & \colhead{(Vega mag)}                       & \colhead{(Gyr)}

}
\startdata
0.165 & 0.15-0.175 & 2.10 & 0.04 & 2.034\\ 
0.358 & 0.3-0.4    & 1.704 &0.03 & 3.923 \\
0.525 & 0.45-0.6   & 1.70 & 0.03 & 5.217\\
0.720 & 0.65-0.78  & 1.48 & 0.03 & 6.432 \\
0.821 & 0.78-0.86  & 1.27 & 0.03 & 6.960\\ 
0.923 & 0.86-0.97  & 1.045 & 0.03& 7.437 \\
1.097 & 1.05-1.15  & 1.1 & 0.03  & 8.139\\
1.246 & 1.2-1.3    & 1.05 & 0.03 & 8.647 \\
1.397 & 1.3-1.5    & 0.68& 0.03   & 9.091\\	 
1.552 & 1.5-1.6    & 0.85& 0.03 & 9.486	\\
1.646 & 1.6-1.7    & 0.59& 0.03 & 9.700\\ 
\enddata
\end{deluxetable}


\begin{deluxetable}{ccccc}
\tabletypesize{\scriptsize}
\tablecaption{Number and Stellar Mass Density of Galaxies with $\log M_*/M_\sun>11$ \label{tbl-3}}
\tablewidth{0pt}
\tablehead{
\colhead{Redshift Interval} & \colhead{$\phi_{red}$} & \colhead{log $\rho_{*_{red}}$} & \colhead{log $\rho_{*_{blue}}$} & \colhead{log $\rho_{*_{all}}$}\\ 
\colhead{}                  & \colhead{($10^{-4}$ Mpc$^{-3}$)} & \colhead{(M$_{\sun}$ Mpc$^{-3}$)}      & \colhead{(M$_{\sun}$ Mpc$^{-3}$)}      & \colhead{(M$_{\sun}$ Mpc$^{-3}$)}

}
\startdata
0.2-0.6    &2.550  &7.627  &6.825  &7.691\\
0.6-0.79   &2.911  &7.671  &6.563  &7.704\\
0.79-0.94  &3.169  &7.765  &6.932  &7.824\\
0.94-1.07  &1.890  &7.508  &7.117  &7.656\\
1.07-1.19  &2.555  &7.648  &7.167  &7.825\\
1.19-1.3   &1.984  &7.538  &7.241  &7.716\\
1.3-1.55   &1.323  &7.438  &7.025  &7.579\\
1.55-1.78  &1.125  &7.337  &7.229  &7.586\\
1.78-2.0   &0.643  &7.112  &6.818  &7.346\\
\enddata
\end{deluxetable}


\end{document}